\documentclass[preprint]{elsarticle}

\usepackage{amsmath,amssymb,amsthm}
\usepackage{latexsym}
\usepackage{graphicx}
\usepackage{float}

\theoremstyle{definition}

\newcommand{\Grd}{\operatorname{\rm Grad}}
\newcommand{\tr}{\operatorname{\rm tr}}

\newcommand{\m}{\mathbf{m}}
\newcommand{\e}{\mathbf{e}}
\newcommand{\F}{\mathbf{F}}
\newcommand{\C}{\mathbf{C}}
\newcommand{\M}{\mathbf{M}}
\newcommand{\I}{\mathbf{I}}
\newcommand{\T}{\mathbf{T}}
\renewcommand{\S}{\mathbf{S}}
\renewcommand{\P}{\mathbf{P}}
\newcommand{\x}{\mathbf{x}}
\newcommand{\X}{\mathbf{X}}

\newcommand{\Lin}{\operatorname{Lin}}
\newcommand{\de}[2]{\frac{\partial #1}{\partial #2}}

\begin{document}

\begin{frontmatter}
\title{Strain-dependent internal parameters in hyperelastic
  biological materials}

\author[a]{G. Giantesio}
\ead{giulia.giantesio@unicatt.it}

\author[a]{A. Musesti}
\ead{alessandro.musesti@unicatt.it}

\address[a]{Dipartimento di Matematica e Fisica ``N. Tartaglia'',\\ Universit\`a
  Cattolica del Sacro Cuore,\\ via dei Musei 41, 25121 Brescia, Italy}
\date{\today}

\begin{abstract}
  The behavior of hyperelastic energies depending on an internal
  parameter, which is a function of the deformation gradient, is
  discussed. As an example, the analysis of two models where the
  parameter describes the activation of a tetanized skeletal muscle
  tissue is presented. In those models, the activation parameter
  depends on the strain and it is shown the importance of considering
  the derivative of the parameter with respect to the strain in order
  to capture the proper stress-strain relations.
\end{abstract}


\end{frontmatter}

\section{Introduction}

The theory of hyperelasticity, where the stress derives from a strain
energy density, is widely used for modeling the nonlinear mechanical
response of many biological materials, see for instance
\cite{Chagnon2015}. Denoting with ${\F}$ the deformation gradient and
with $W({\F})$ the strain energy density of a homogeneous hyperelastic
material, one can express the first Piola-Kirchhoff stress tensor
field $\P$ as
\[
{\P}=\frac{\partial W}{\partial {\F}}(\F).
\]
Hence, in the choice of constitutive prescriptions only the scalar
quantity $W$ needs to be described, which is much simpler than modeling a
tensor such as the stress.

Furthermore, for some complex materials an {\em internal parameter}
$\gamma$ can be introduced \cite{LazOgd,OgdRox}, in order to account
for microstructural changes, and the energy density becomes a function
both of $\F$ and $\gamma$. For simplicity, we assume that $\gamma$ is
scalar valued, although similar considerations can be carried on also
when the parameter is vector or tensor valued.

In some situations it is customary to model the internal parameter as
a function of the deformation gradient, that is
$\gamma=\gamma(\F)$. This happens, for instance, in the description of a
  tetanized skeletal muscle tissue, where the strength developed by
  the sarcomeres depends on the overlap between the actin and myosin
  chains, and the overlap changes with the deformation. In
  \cite{RajWin} a similar approach is applied to quite a different
  situation, namely the description of strain-induced microstructural
  changes in the context of plasticity. In that case, however, these
  changes are permanent and the hyperelastic behavior is lost. 
  
  In general, if one persists in the context of hyperelasticity, the
  dependence of the parameter $\gamma$ on the deformation gradient has to be taken
  into account in the expression of the stress:
\[
{\P}=\frac{\partial W}{\partial {\F}}(\F,\gamma(\F))+
\frac{\partial W}{\partial {\gamma}}(\F,\gamma(\F))
\frac{\partial \gamma}{\partial {\F}}(\F).
\]
The previous calculation has often been overlooked in the literature,
especially in the modeling of active biological tissues, and the
simpler expression ${\P}=\frac{\partial W}{\partial {\F}}(\F,\gamma)$
has been used instead of the former (see for
instance~\cite{Bol11,ebi,Hernandez13,WIMJ14,noi}).

In this paper we want to emphasize the correct expression of the
stress when the internal parameter depends on the strain and one
persists in the context of hyperelasticity. Moreover, we propose two
reliable models for the tetanized skeletal muscle tissue. The issue
has been widely studied in the recent
literature~\cite{ebi,Stahland2013,Hernandez13,WIMJ14,thomas,noi}; the
new approaches here proposed use the same passive strain energy
function, which has been introduced in~\cite{ebi}, while the
activation is described in two different ways. Notice that the
proposed models consider the behavior of a muscle when the activation
is at its maximum: the amount of activation cannot be voluntarily
controlled, but it is an experimental datum which depends only on the
strain.

\section{Theoretical framework} 
\label{theory}


In Continuum Mechanics the motion of a body is described by an
invertible smooth map from a bounded subset
$\Omega\subset\mathbb{R}^3$ into $\mathbb{R}^3$: the function
${\x} = \chi({\X}, t)$ associates every point $\X$ in the reference
configuration $\Omega$ with its current placement $\x$.  The
\emph{deformation gradient}
\begin{equation}
{\F} = \Grd {\chi}, \ \ F_{ij}=\de{x_i}{X_j}, i, j= 1,2,3 \nonumber
\end{equation}
belongs to the space of linear operators with strictly positive
determinant ($\Lin^+$).

The tensional state in a continuum is described in material
coordinates by the first Piola-Kirchhoff (or nominal) stress tensor
${\P}$, which is related to the Cauchy
stress tensor ${\T}$ and the second Piola-Kirchhoff stress tensor
${\S}$ by
\begin{equation}
{\T} = J^{-1} {\P}{\F}^{T}, \quad
{\S} = {\F}^{-1} {\P} = J {\F}^{-1} {\T}{\F}^{-T}, \qquad
J = \det\F. \nonumber
\end{equation}

In Elasticity, a relation between the stress and the deformation
gradient $\F$ is assumed, so that $\P$ can be expressed as a function
of $\F$. If the material is supposed to be \emph{hyperelastic}, there
exists a strain energy density function
$W : \Omega\times\Lin^+ \rightarrow \mathbb{R}$ such that
\begin{equation} \label{PiolaF}
{\P}(\X,\F)=\frac{\partial W}{\partial {\F}}(\X,\F).
\end{equation}
Hence the behavior of the elastic body is described by a hyperelastic
strain energy function
\begin{equation}
 \int_{\Omega} W(\X,\F)dV. \nonumber
\end{equation}
For the sake of simplicity, from now on we will assume that the
material is \emph{homogeneous}, so that the explicit dependence of $W$
on $\X$ can be dropped.

The hyperelastic model can be useful in describing the elastic
behavior of many biological tissues. Moreover, in order to account for
biological phenomena such as, for instance, activation and growth, a parameter is
introduced in the function $W$ which keeps into account of
(micro)structural changes of the material, usually related to some
chemical reactions. We denote by $\gamma$ such a parameter which
describes the internal state of the material. Then the elastic energy
density writes
\begin{equation}
\label{eq:gamma}
W=W({\F},\gamma).
\end{equation}
Even if $\gamma$ can be a vector or a tensor quantity
(see~\cite[Sect.\ 2]{LazOgd}), in the sequel we will assume that
it is a scalar parameter.

We are interested in some applications, such as the description of
a tetanized skeletal muscle tissue, where it is necessary to relate the
parameter $\gamma$ to the deformation gradient $\F$, that is
$\gamma=\gamma({\F})$; in that case, the nominal stress tensor is given
by
\begin{equation}
  {\P}=\de{W}{{\F}}({\F},\gamma({\F}))+\de{W}{{\gamma}}({\F},\gamma({\F}))
\de{\gamma}{\F}({\F}).\label{Piolagammanoncost}
\end{equation}
However, in the literature, the term $\de{W}{{\gamma}}\de{\gamma}{\F}$ is often
neglected, so that the stress is not the derivative of the strain
energy density with respect to the deformation gradient. Our aim is to show
that this term is important in order to describe the tensional state
of the hyperelastic material.

We remark that in some papers, such as \cite{LazOgd,Horgan,Pae12}, it
is assumed that $\de{W}{{\gamma}}=0$, so
that~\eqref{Piolagammanoncost} reduces to the simpler expression
${\P}=\de{W}{{\F}}({\F},\gamma)$; from that assumption, a relation
between $\gamma$ and $\F$ is deduced. Nevertheless, there are many
situations in which the relation between $\gamma$ and $\F$ comes from
biological data and does not satisfy the constraint
$\de{W}{{\gamma}}=0$.

In the next section we will give two
examples in which the parameter $\gamma$ describes the activation state
of a tetanized muscle and the term
$\de{W}{{\gamma}}\de{\gamma}{\F}$ cannot be
neglected.

\section{Examples related to the activation of skeletal muscle tissue}

One of the main features of the muscle tissue is its ability of
activating through a chemical reaction between actin and myosin
filaments, which induces a contraction of the muscle fibers.
The two filaments form the basic motor unit of the muscle, that is the
\emph{sarcomere}. A typical stress-stretch curve of a sarcomere
reveals that the amount of activation is a function of the
stretch~\cite[Chapter 4]{Carol}. For this reason in the literature the
activation of a skeletal muscle usually depends on the deformation.

\begin{figure}[htbp]
        \includegraphics[width=0.80\textwidth]{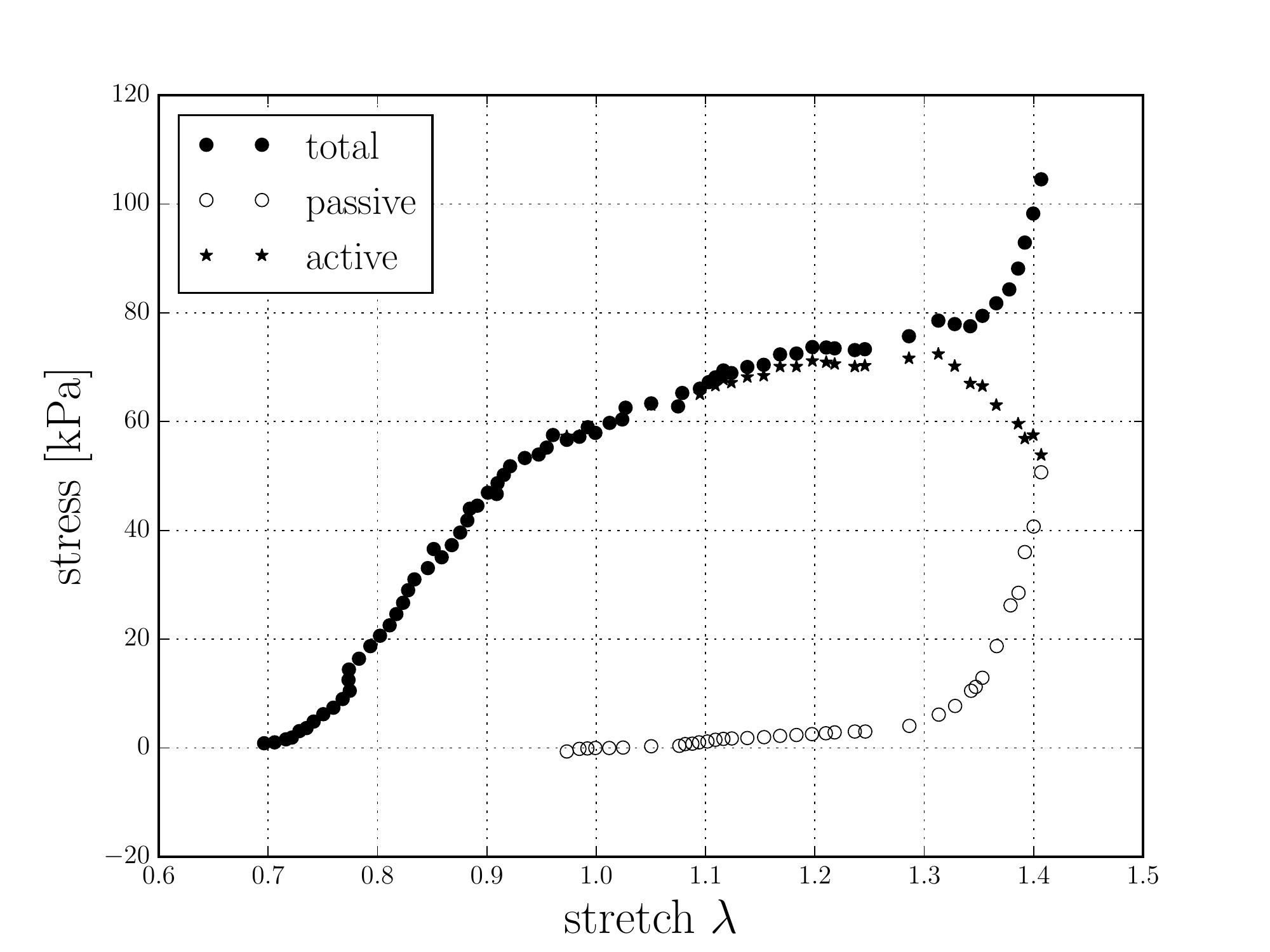}
        \caption{Stress-stretch relationship reported in~\cite{hawkins}.}
     \label{hb}
\end{figure}

The experiments {\em in vivo} are often performed in two
steps:
\begin{enumerate}
\item first, the stress-stretch relation is measured without activation,
  obtaining the so called \emph{passive curve};
\item second, the muscle is kept in a tetanized state by
  an electrical stimulus and the \emph{total stress-stretch curve} is
  plotted.
\end{enumerate}
The so called \emph{active curve}, which describes the amount of
stress due to activation, is obtained by taking the difference of the
two previous curves. In Figure~\ref{hb} we show the three stress-stretch curves
obtained in a celebrated experiment \emph{in vivo} by Hawkins and Bey~\cite{hawkins}
for a tetanized \emph{tibialis anterior} of a rat.

In order to model the activation of muscle tissue, a parameter can be
introduced in the hyperelastic energy, such as in
equation~\eqref{eq:gamma}.  We will now present two different ways of
including the activation parameter in the same strain energy function; both
of them aim to reproduce the active data of Figure~\ref{hb}.

\null

Following the model by Ehret, B{\"{o}}l and Itskov~\cite{ebi}, we
describe the passive behavior of the skeletal muscle tissue as a
transversely isotropic hyperelastic incompressible material. By
frame-indifference, the passive strain energy density depends only on
the right Cauchy-Green tensor $\C={\F}^T{\F}$ and is assumed to be
\begin{equation}
W_p({\C})=\frac{\mu}{4}\left\{\frac{1}{\alpha}\left[e^{\alpha({I}_p-1)}-1\right]+
\frac{1}{\beta}\left[e^{\beta(K_p-1)}-1\right]\right\}, 
\quad \det\C=1,\label{Webipassivevero}
\end{equation}
where 
\begin{equation}
{I}_p=\frac{w_0}{3}\tr({\C})+(1-w_0)\tr({\C}{\M}), \ \
K_p=\frac{w_0}{3}\tr({{\C}^{-1}})+(1-w_0)\tr({\C}^{-1}\M). \nonumber
\end{equation}
The tensor $\M=\m\otimes\m$ is called {\em structural tensor}, $\m$
being the orientation of the fibers.
The incompressibility constraint $\det\C=1$ is due to the large amount
of water contained in the skeletal muscle tissue. The material
parameters $\alpha, \beta, w_0$ and $\mu$ are related to the passive
proprieties of the muscle (in particular, $w_0$ measures the amount of
anisotropy of the material).

In the following, we want to modify the passive model in order to
account for the activation. Hence it is useful to introduce a
constitutive assumption for $P_{act}$\,, which is the active part of
the stress given in Figure~\ref{hb}. As in~\cite{ebi}, we
assume that
\begin{equation}\label{pact}
 P_{act}(\lambda) =
\left\{
\begin{aligned}
&P_{opt}\frac{\lambda_{min}-\lambda}{\lambda_{min}-\lambda_{opt}}e^{\frac{(2\lambda_{min}-\lambda-\lambda_{opt})(\lambda-\lambda_{opt})}{2(\lambda_{min}-\lambda_{opt})^2}}
& \text{if $\lambda>\lambda_{min}$},
\\[1ex]
&0 & \text{otherwise}.
\end{aligned}
\right. 
\end{equation}

The values of the material parameters, which have been obtained
in~\cite{ebi} by least squares optimization using the experimental
data, are given in Table~\ref{parpass}.  Notice that the
value of $P_{opt}$ takes into account some information at the
mesoscale level, such as the number of activated motor units and the
interstimulus interval; in our model, $P_{opt}$ describes the maximal
activation, which is reached in the tetanized case.
\begin{table}[h]
\caption{Material parameters.}
\label{parpass}
\begin{tabular}{p{1.4cm}p{1cm}p{1cm}p{1.1cm}p{1.4cm}p{1.4cm}p{1.6cm}}
\hline\noalign{\smallskip}
$\mu$ [kPa] & $\alpha$ [-] & $\beta$ [-] & $w_0$ [-]
& $\lambda_{min}$ [-] & $\lambda_{opt}$ [-] & $P_{opt}$ [kPa]\\
\hline\noalign{\smallskip}
0.1599 & 19.69 & 1.190 & 0.7388 
& 0.682 & 1.192 &73.52 \\
\hline\noalign{\smallskip}
\end{tabular}
\end{table}

\subsection{First activation model: modification of the generalized
  invariant $I_p$ }
\label{sectebi}

A simple way to model the activation is proposed in~\cite{ebi}. There,
a parameter $\gamma\geq 0$ is introduced which adjusts the generalized
invariant $I_p$:
\[
\widetilde{I}=I_p+\gamma\tr({\C}{\M})=\frac{w_0}{3}\tr({\C})+(1-w_0+\gamma)\tr({\C}{\M}).
\]
In fact, the parameter $\gamma$ increases the amount of elastic energy
during an elongation in the direction of the fibers.
Then the total strain energy density becomes
\begin{equation}
W({\C},\gamma)=\frac{\mu}{4}\left\{\frac{1}{\alpha}\left[e^{\alpha(\widetilde{I}-1)}-1\right]+
\frac{1}{\beta}\left[e^{\beta(K_p-1)}-1\right]\right\},
\quad\det\C=1.\label{Webiactive}
\end{equation}

The stress evaluated on a uniaxial deformation along the fibers
should fit the total stress of the experimental data reported
in~Figure~\ref{hb}; it turns out that $\gamma$ cannot be constant and
it has to be considered as a function of the deformation.
In order to choose a suitable model for the
active parameter $\gamma(\F)$, let us consider a uniaxial deformation of
stretch $\lambda$ along the fiber direction $\m=\e_1$ in the
transversely isotropic incompressible case:
\[
\F=\begin{pmatrix}
\lambda & 0 & 0\\
0 & \frac{1}{\sqrt{\lambda}} & 0\\
0 & 0 & \frac{1}{\sqrt{\lambda}}
\end{pmatrix}.
\]
Then the elastic energy density writes
\begin{equation}
\label{energyebi}
W(\lambda,\gamma(\lambda))=\frac{\mu}{4}\left\{\frac 1 \alpha\left[e^{\alpha(\widetilde{I}(\lambda, \gamma(\lambda))-1)}-1\right]+
\frac 1 \beta\left[e^{\beta(K_p(\lambda)-1)}-1\right]\right\}
\end{equation}
with
\begin{align}
&\widetilde{I}(\lambda,
\gamma(\lambda))=\frac{w_0}{3}\left({\lambda^2}+\frac{2}{\lambda}\right)+(1-w_0+\gamma(\lambda)){\lambda^2}=I_p(\lambda)+\gamma(\lambda)\lambda^2,
\label{Itilde}\\[1ex]
&K_p(\lambda)=\frac{w_0}{3}\left(\frac{1}{\lambda^2}+2\lambda\right)+\frac{1-w_0}{\lambda^2}.\label{Kp}
\end{align}
Using~\eqref{Piolagammanoncost}, the total stress $P_{tot}$ and
the passive stress $P_{pas}$ are given by
\begin{align}
\label{rightstress}
&P_{tot}(\lambda,\gamma(\lambda))=\de{W}{\lambda}(\lambda,\gamma(\lambda))+
\de{W}{\gamma}(\lambda,\gamma(\lambda))\gamma^\prime(\lambda),\\[1ex]
&{P}_{pas}(\lambda)=\de{W}{\lambda}(\lambda,0).
\end{align}
Hence, in order to fit the experimental data of the total stress, $\gamma$
has to solve the equation
\begin{equation}
{P}_{tot}(\lambda, \gamma(\lambda))= 
{P}_{pas}(\lambda)+{P}_{act}(\lambda), \label{eqgammaimplicita}
\end{equation}
which leads to the Cauchy problem
\begin{equation}\label{problemadiffebi}
\left\{
\begin{aligned}
&\de{W}{\lambda}(\lambda,\gamma(\lambda))+
\de{W}{\gamma}(\lambda,\gamma(\lambda))\gamma^\prime(\lambda)=
P_{pas}(\lambda)+P_{act}(\lambda),
\\[1ex]
&\gamma(\lambda_{min})=0.
\end{aligned}
\right. 
\end{equation}
The condition $\gamma(\lambda_{min})=0$ comes from the expression of $P_{act}$ in \eqref{pact}, \emph{i.e.}\ we assume that the activation begins after $\lambda_{min}$.

\begin{figure}[h!]
\includegraphics[width=.8\textwidth]{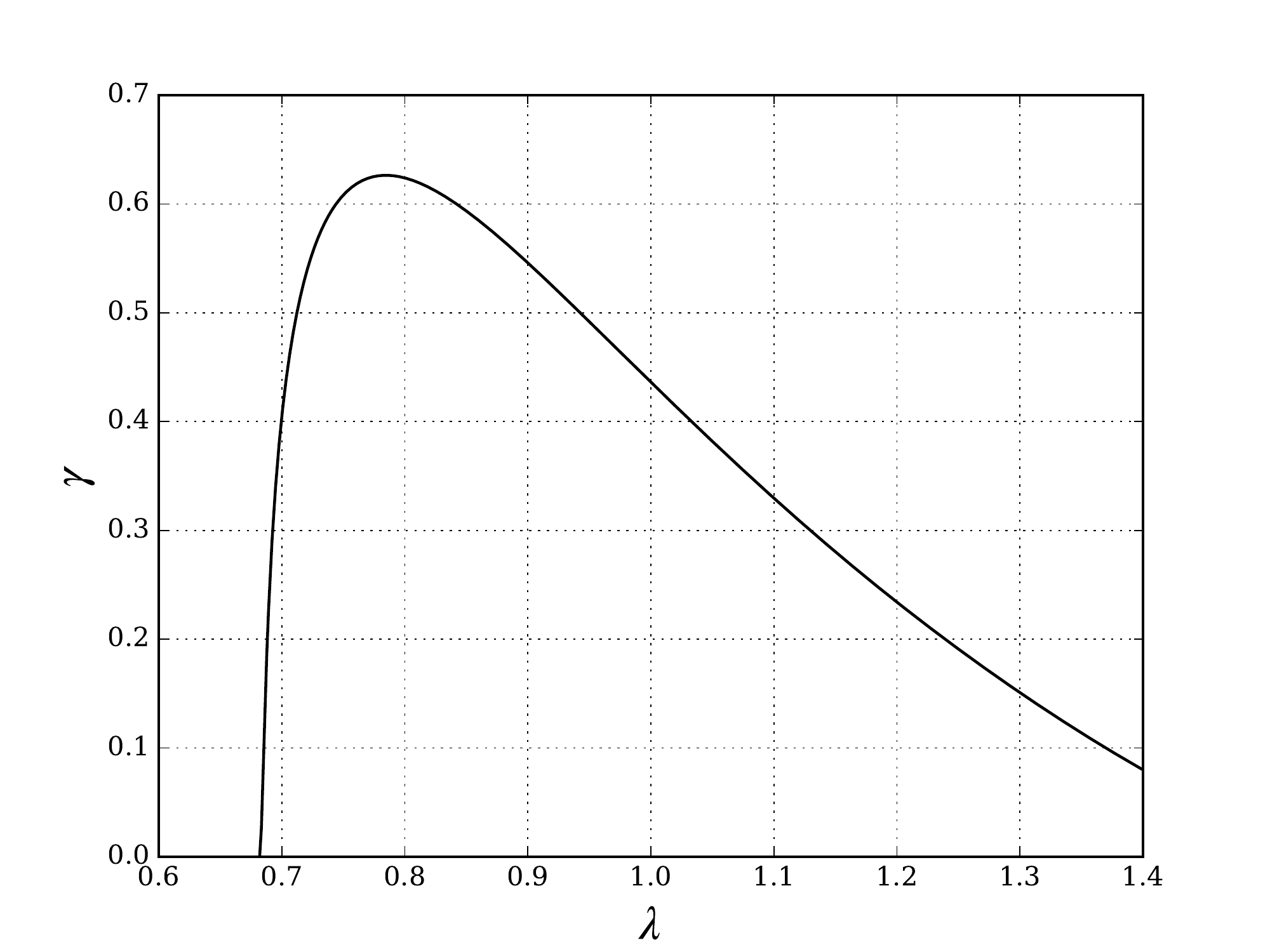}
\caption{The behavior of the activation parameter $\gamma(\lambda)$
  give in~\eqref{soluzionewagiusta}.}
\label{figgamma}
\end{figure}

We note that~\eqref{problemadiffebi} can be explicitly integrated, so
that $\gamma$ can be found by solving the equation
\begin{equation}\label{problemawaintegrato}
W(\lambda, \gamma(\lambda))= W(\lambda, 0) +S_{act}(\lambda),
\end{equation}
where
\begin{equation}\label{sact}
 S_{act}(\lambda) =
\left\{
\begin{aligned}
&\int_{\lambda_{min}}^{\lambda} P_{act}(\xi)d\xi
& \text{if $\lambda>\lambda_{min}$},
\\[1ex]
&0 & \text{otherwise},
\end{aligned}
\right. 
\end{equation}
and 
\begin{align}
\int_{\lambda_{min}}^{\lambda} P_{act}(\xi)d\xi =P_{opt}(\lambda_{min}-\lambda_{opt})\left[e^{\frac{(2\lambda_{min}-\lambda-\lambda_{opt})(\lambda-\lambda_{opt})}{2(\lambda_{min}-\lambda_{opt})^2}} - e^{\frac{1}{2}}\right].\label{intsact}
\end{align}
The solution of equation~\eqref{problemawaintegrato} is given by 
\begin{equation}\label{soluzionewagiusta}
\gamma(\lambda)=\frac{1}{\alpha \lambda^2} \ln{\left\{ 1+
    \frac{4\alpha}{\mu}
    S_{act}(\lambda)e^{\left[1-\frac{w_0}{3}\left(\lambda^2+\frac{2}{\lambda}\right)-(1-w_0)\lambda^2\right]}
  \right\}}
\end{equation}
and is represented in Figure~\ref{figgamma}. 

In the Appendix we will briefly comment the function
$\gamma(\lambda)$ given in~\cite{ebi}, which is different
from~\eqref{soluzionewagiusta}. Indeed, in the computation of
$P_{tot}$ in~\eqref{rightstress} we took into account that $\gamma$ has
to depend on the stretch $\lambda$, while in~\cite{ebi} the derivative
of the energy is computed as if $\gamma$ be a constant.

\null

It is interesting to perform a numerical simulation for a general
deformation. In that case, the stretch of the fibers $\lambda$ is given
by the quantity
\[
I_4=\sqrt{\tr({\C}{\M})},
\]
so that the energy density of the hyperelastic active skeletal muscle tissue is
\[
W(\C,\gamma(I_4)),
\]
where $W$ and $\gamma$ are given in~\eqref{Webiactive}
and~\eqref{soluzionewagiusta}, respectively. Taking into account that
$\de{I_4}{\C}=\M$, the expression of the stress becomes
\[
{\P}=2\F\de{W}{\C}(\C,\gamma(I_4))+
2\de{W}{\gamma}(\C,\gamma(I_4))\,\de{\gamma}{I_4}(I_4)\,\F\M-p\F^{-T},
\]
where $p$ is a Lagrange multiplier associated with the hydrostatic
pressure which results from the incompressibility constraint $\det\C=1$.

\begin{figure}[h!]
\includegraphics[width=.8\textwidth]{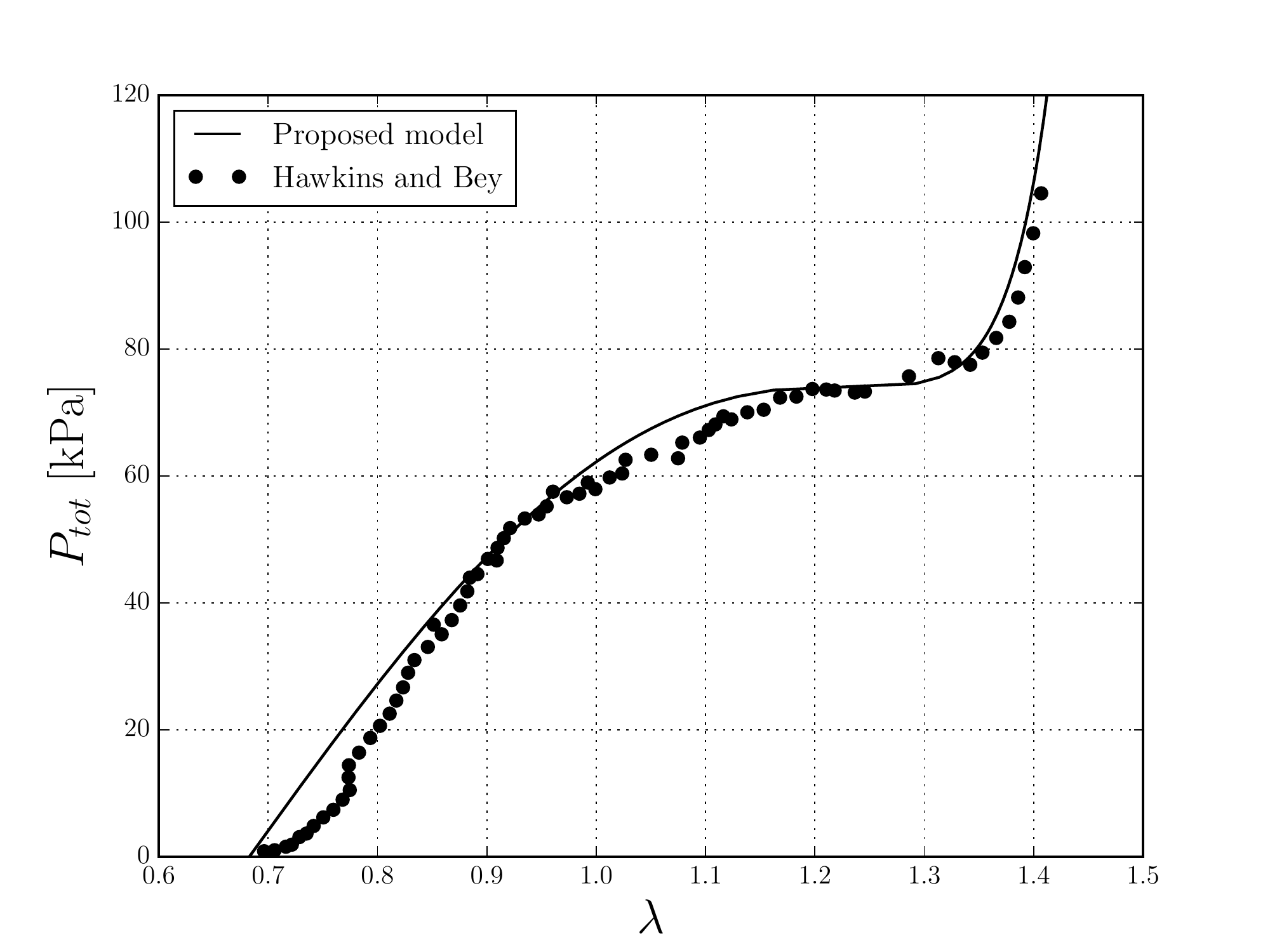}
\caption{Stress-stretch relation obtained using~\eqref{soluzionewagiusta}.}
\label{figsforzogiusto}
\end{figure}

We simulate numerically a uniaxial deformation of a
cylindrical slab of tissue, along the axis of the
cylinder. We assume radial symmetry, so that the computational mesh
reduces to a rectangle. 
Concerning the boundary conditions, the bases
of the cylinder are kept perpendicular to the axial
direction. Moreover, a basis of the cylinder is fixed and a load is
applied to the other. The finite element simulation is performed using
the open source project FEniCS~\cite{fenics}, a collection of
numerical software, supported by a set of novel algorithms and
techniques, aimed at the automated solution of differential equations.
In Figure~\ref{figsforzogiusto} we show the stress-stretch relation
obtained numerically, which fits the experimental data quite well.

\subsection{Second activation model: the active strain approach}
\label{sectas}
\begin{figure}[htbp]
        \includegraphics[width=0.70\textwidth]{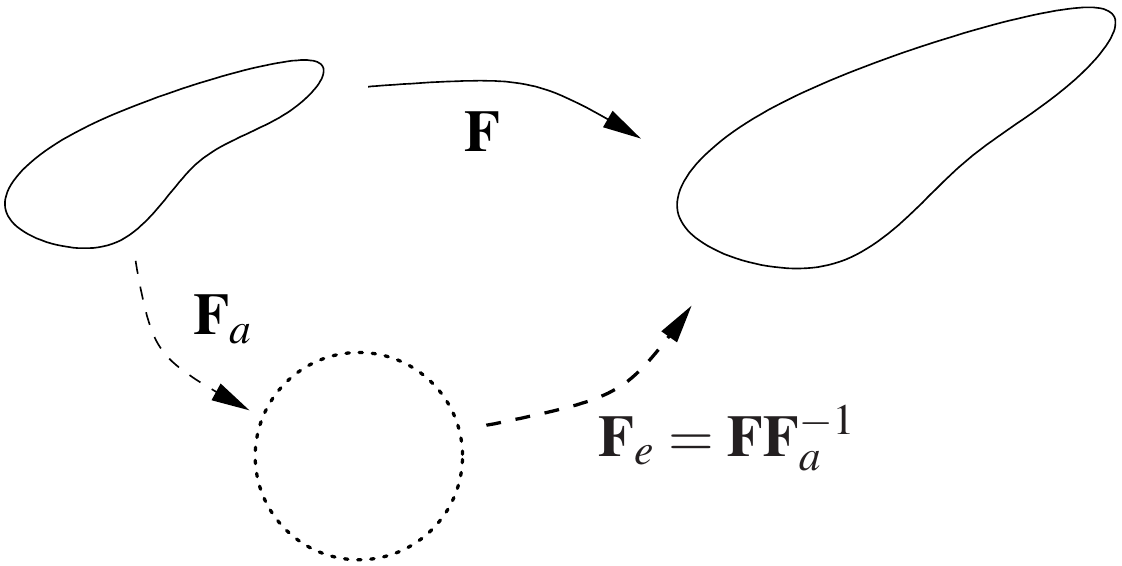}
   \caption{Pictorial view of the Kr\"oner-Lee decomposition in the
     active strain approach.} 
     \label{figactivestrainapproach}
\end{figure}

A more interesting way of modeling activation through a parameter is
by means of active strain. Actually, the activation model considered
in the previous section is halfway between active strain and active
stress, which are the two main paths followed in the literature (for a
review see \cite{AmbPez}).  The so called \emph{active stress} method
adds an extra contribution to the stress, accounting for activation
(see for example \cite{martins, blemker, thomas}). On the other hand,
the \emph{active strain} approach assumes that only a part of the
deformation gradient is responsible for storing the elastic energy,
although the form of the strain energy function does not change.
Similarly to the Kr\"oner-Lee decomposition of finite
plasticity~\cite{DiCarlo02}, the deformation gradient is written as
${\F}={\F}_e{\F}_a$, where ${\F}_e$ is the elastic part and ${\F}_a$
describes the active contribution (see
Figure~\ref{figactivestrainapproach}). The tensor $\F_a$, which needs
not be the gradient of some displacement, has a clear biological
interpretation, since it is related to the sliding movement of the
filaments in the sarcomeres, which is the main mechanism of
contraction at the mesoscale. This method was first proposed by Taber
and Perucchio \cite{taber} for the activation of the cardiac tissue,
and it is detailed in \cite{NarTer} for soft living tissues. In
\cite{noi} the active strain approach has been specifically applied to
skeletal muscle tissue.

Given a passive strain energy density $W_p$, the hyperelastic energy
density of the activated material is given by
\[
{W}({\C},\F_a) = (\det{\F}_a)W_p({\C}_e),\quad
\C_e={\F}^{-T}_a{\C}{\F}_a^{-1},
\]
where the active strain
${\F}_a$ has to be constitutively prescribed as a function of the
deformation gradient.

Let us consider the same passive strain energy
density function given in \eqref{Webipassivevero} and choose the
incompressible activation as
\begin{equation}
 {\F}_a(I_4)=(1-\gamma(I_4)){\m}\otimes{\m}+\frac{1}{\sqrt{1-\gamma(I_4)}}({\I}-{\m}\otimes{\m}),
 \quad I_4=\sqrt{\tr(\C\M)},\label{Fa} 
\end{equation}
where $0\leq\gamma <1$ represents the relative contraction of
activated fibers ($\gamma=0$ meaning no contraction). Since
$\det\F_a=1$, the modified strain energy density becomes
\begin{equation}
{W}({\C},\gamma(I_4))=W_p({\C}_e)=\frac{\mu}{4}\left\{\frac{1}{\alpha}\left[e^{\alpha(I_e-1)}-1\right]+\frac{1}{\beta}\left[e^{\beta(K_e-1)}-1\right]
\right\},
\label{Wattiva}
\end{equation}
where $\det\C=1$ and
\begin{align*}
&I_e=\frac{w_0}{3}\tr({\C}_e)+(1-w_0)\tr({\C}_e{\M}),\\
&K_e=\frac{w_0}{3}\tr({{{\C}_e}^{-1}})+(1-w_0)\tr({{\C}_e}^{-1}{\M}).
\end{align*}

As in Section \ref{sectebi}, the parameter $\gamma$ has to be
calibrated in order to obtain the active curve given in
\eqref{pact}. Let us consider a uniaxial incompressible tension along
the fiber direction ${\m}={\e}_1$, so that
\[
\C_e=
\begin{pmatrix}
\frac{\lambda^2}{(1-\gamma(\lambda))^2} & 0 & 0\\
0 & \frac{1-\gamma(\lambda)}{\lambda} & 0\\
0 & 0 & \frac{1-\gamma(\lambda)}{\lambda}
\end{pmatrix}.
\]
Then the Cauchy problem~\eqref{problemadiffebi} reduces again to equation
\eqref{problemawaintegrato} which, after some trivial simplifications, writes
\begin{equation}\label{problemaebiactivestrain}
\frac{1}{\alpha}e^{\alpha({I}_e(\lambda,\gamma(\lambda))-1)}+
\frac{1}{\beta}e^{\beta(K_e(\lambda,\gamma(\lambda))-1)}=
\frac{1}{\alpha}e^{\alpha({I}_p(\lambda)-1)}+
\frac{1}{\beta}e^{\beta(K_p(\lambda)-1)}+\frac{4}{\mu}S_{act}(\lambda),
\end{equation}
where
\begin{align*}
&I_e(\lambda,\gamma(\lambda))=\frac{w_0}{3}\left[\frac{\lambda^2}{(1-\gamma(\lambda))^2}+\frac{2(1-\gamma(\lambda))}{\lambda}\right]+(1-w_0)\frac{\lambda^2}{(1-\gamma(\lambda))^2},\\[1ex]
&K_e(\lambda,\gamma(\lambda))=\frac{w_0}{3}\left[\frac{(1-\gamma(\lambda))^2}{\lambda^2}+2\frac{\lambda}{1-\gamma(\lambda)}\right]+(1-w_0)\frac{(1-\gamma(\lambda))^2}{\lambda^2},\\[1ex]
&I_p(\lambda)=I_e(\lambda,0), \quad K_p(\lambda)=K_e(\lambda,0).
\end{align*}

\begin{figure}[htbp]
        \includegraphics[width=0.80\textwidth]{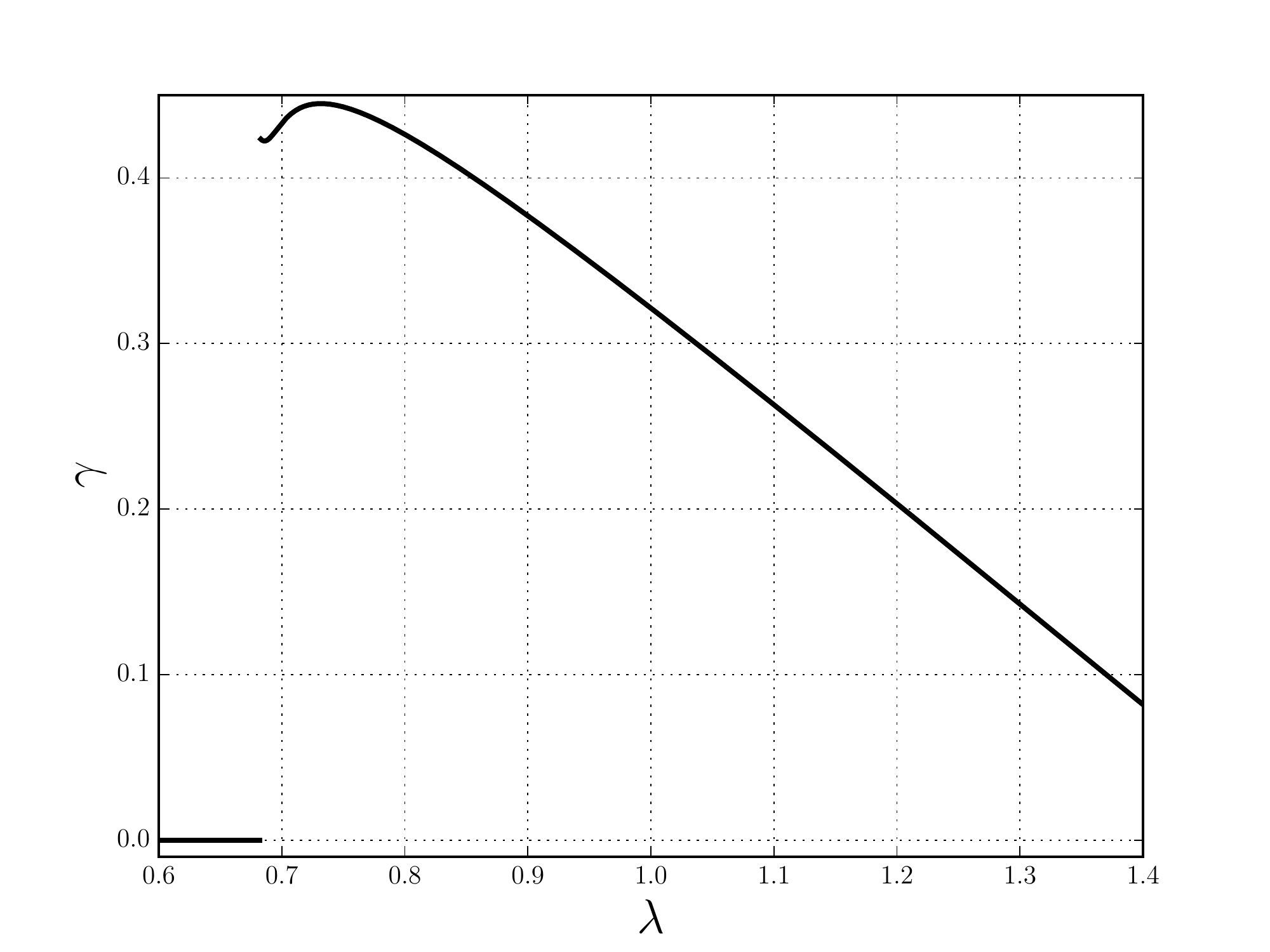}
   \caption{Numerical solution of equation~\eqref{problemaebiactivestrain}.} 
     \label{figgammabise}
\end{figure}

Differently from the previous model,
equation~\eqref{problemaebiactivestrain} cannot be explicitly solved,
because the dependence of $W$ on $\gamma$ is much more complicated
than before. However, one can employ standard numerical methods in
order to find the solution. Figure~\ref{figgammabise}, which is
obtained by a bisection method, shows the solution $\gamma$ as a
function of $\lambda$.
The behavior of the solution is quite similar to that shown in
Figure~\ref{figgamma}, in the sense that it has a maximum point
and then decreases to 0 for large $\lambda$. However, in this case
$\gamma(\lambda)$ is discontinuous in $\lambda_{min}=0.682$. 

\begin{figure}[h!]
\includegraphics[width=.8\textwidth]{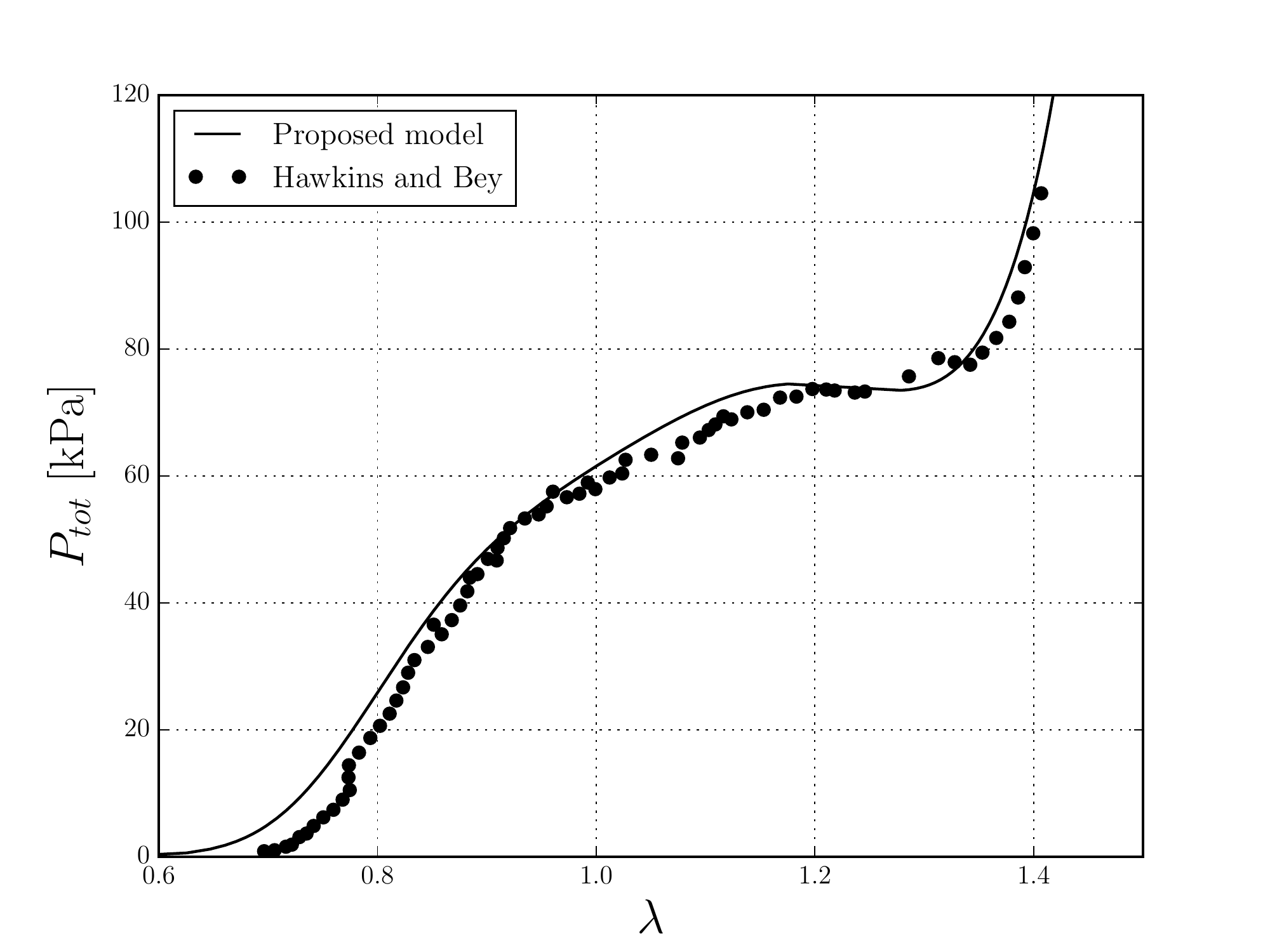}
\caption{Stress-stretch relation obtained using the active strain approach.}
\label{figsforzoactivestrain}
\end{figure}

In order to perform a finite element simulation, it is convenient to
have an explicit function $\gamma(\lambda)$. A possible strategy is to
choose a suitable form for $\gamma$ and to fit this expression to the
data. Using a polynomial of degree 5 and running the simulation as
in the previous section, we obtain Figure
\ref{figsforzoactivestrain}. As one can see, the numerical results
describe quite well the experimental data.

\section{Concluding remarks}
We analyzed the behavior of hyperelastic materials depending on a
strain-dependent internal parameter $\gamma$. The parameter is
introduced in the strain energy density function in order to capture
some real phenomena such as growth and activation of biological
tissues. In the fitting of experimental data, it is 
necessary to relate the internal parameter with the deformation
gradient: in that case, if the
material is supposed to be hyperelastic, a further term appears in the
stress tensor, due to the derivative of $\gamma$ with respect to the
deformation gradient.
The present paper underlines the importance of
considering such a term in order to get the proper behavior of the
stress.

To this aim, we focused our attention on active skeletal muscle
tissue, where it is often useful to consider an internal parameter
describing the activation state of the muscle in its tetanized state.
Starting from~\cite{ebi,noi}, two different ways of including $\gamma$
in the hyperelastic energy are presented. First, activation is
described following the approach used in \cite{ebi} and the corresponding
hyperelastic energy is found. Second, the active strain approach
studied in~\cite{noi} is used. In both cases, $\gamma$ is given as the
solution of an equation which relates the passive and active strain
energy functions.  We remark that the first approach is not customary
in the literature, while the active strain method seems to be closer
to the biological meaning of activation.  However, the simplicity of
the first method allows to find an explicit function for $\gamma$,
differently from the active strain case.

Notice that, once the dependence of $\gamma$ on the deformation has been
taken into account, the hyperelastic energy of the active material can
lose some of the mathematical properties of the passive one: for
instance, if the latter is polyconvex, as in the case
of~\eqref{Webipassivevero}, the former can even be not rank-1
convex. A future development will be a study of the properties of the
active material depending on the form of the activation function
$\gamma$ and the passive energy.

\appendix 

\section{Comparison with the activation parameter in~\cite{ebi}}

The aim of this Appendix is to point out the differences between the
activation parameter $\gamma(\lambda)$ computed in
Section~\ref{sectebi} and the results of~\cite[Sect. 3.1]{ebi}.

Let us consider again a uniaxial deformation along the fibers and the
elastic energy~\eqref{energyebi}. If one computes the derivative of the energy as if
$\gamma$ were a constant, it follows that
\begin{equation}
\label{Ptotebi}
{P}_{tot}(\lambda, \gamma)= \de{W}{\lambda}(\lambda, \gamma)= \frac{\mu}{4}\left[\de{\widetilde{I}}{\lambda}(\lambda, \gamma)\, e^{\alpha(\widetilde{I}(\lambda, \gamma)-1)}+
\de{K_p}{\lambda}(\lambda)\,e^{\beta(K_p(\lambda)-1)}\right],
\end{equation}
where $\widetilde{I}$ and $K_p$ are given in~\eqref{Itilde}-\eqref{Kp} and
\begin{align*}
&\de{\widetilde{I}}{\lambda}(\lambda, \gamma)=2\frac{w_0}{3}\left({\lambda}-\frac{1}
    {\lambda^2}\right)+2(1-w_0+\gamma){\lambda}=I_p^\prime(\lambda)+2\gamma\lambda,\\[1ex]
&\de{K_p}{\lambda}(\lambda)=2\frac{w_0}{3}\left(-\frac{1}{\lambda^3}+1\right)
-2(1-w_0)\frac{1}{\lambda^3}. 
\end{align*}
The passive part of the stress is given by
${P}_{pas}(\lambda)={P}_{tot}(\lambda, 0)$
and, in order to fit the experimental data of the total stress, $\gamma$
has to solve the equation
\begin{equation*}
{P}_{tot}(\lambda, \gamma(\lambda))= 
{P}_{pas}(\lambda)+{P}_{act}(\lambda),
\end{equation*}
where $P_{act}$ is again given by~\eqref{pact}.
Such an equation is not so
easy to solve in $\gamma$, since it is of the form
\[
\Big(I'_p(\lambda)+2\lambda\gamma\Big)e^{\alpha\lambda^2\gamma}=f(\lambda),
\]
where $f(\lambda)$ is a given function of $\lambda$.
However, after some calculations the solution writes
\begin{equation}\label{waebi}
 \gamma(\lambda) =
\left\{
\begin{aligned}
&\frac{W_0(x^*(\lambda))}{\alpha\lambda^2}-\frac{I_p^{\prime}(\lambda)}{2\lambda} && \text{if $\lambda>\lambda_{min}$},
\\[1ex]
&0
&& \text{otherwise},
\end{aligned}
\right. 
\end{equation}
where $W_0$ is the Lambert W-function, which is the inverse of
$xe^x$,
 and
\begin{align}
&x^*(\lambda)=P_{act}(\lambda)\frac{2\alpha\lambda}{\mu}
e^{\frac{\alpha}{2}(2-2I_p(\lambda)+\lambda I_p^{\prime}(\lambda))}+\frac{\alpha}{2}\lambda I_p^{\prime}(\lambda)e^{\frac{\alpha}{2}\lambda I_p^{\prime}(\lambda)}.
\nonumber
\end{align}
As one can see, even if the internal parameter $\gamma$ is a function
of $\lambda$, the stress~\eqref{Ptotebi} has been computed assuming
that $\gamma$ is independent of the strain. In particular, the
stress~\eqref{Ptotebi} does not come from the hyperelastic energy
$W(\lambda,\gamma(\lambda))$. Moreover, from a mathematical viewpoint
the expression of $\gamma$ in~\eqref{soluzionewagiusta} is much
simpler than the one given in~\eqref{waebi}.

\begin{figure}[h!]
\includegraphics[width=.8\textwidth]{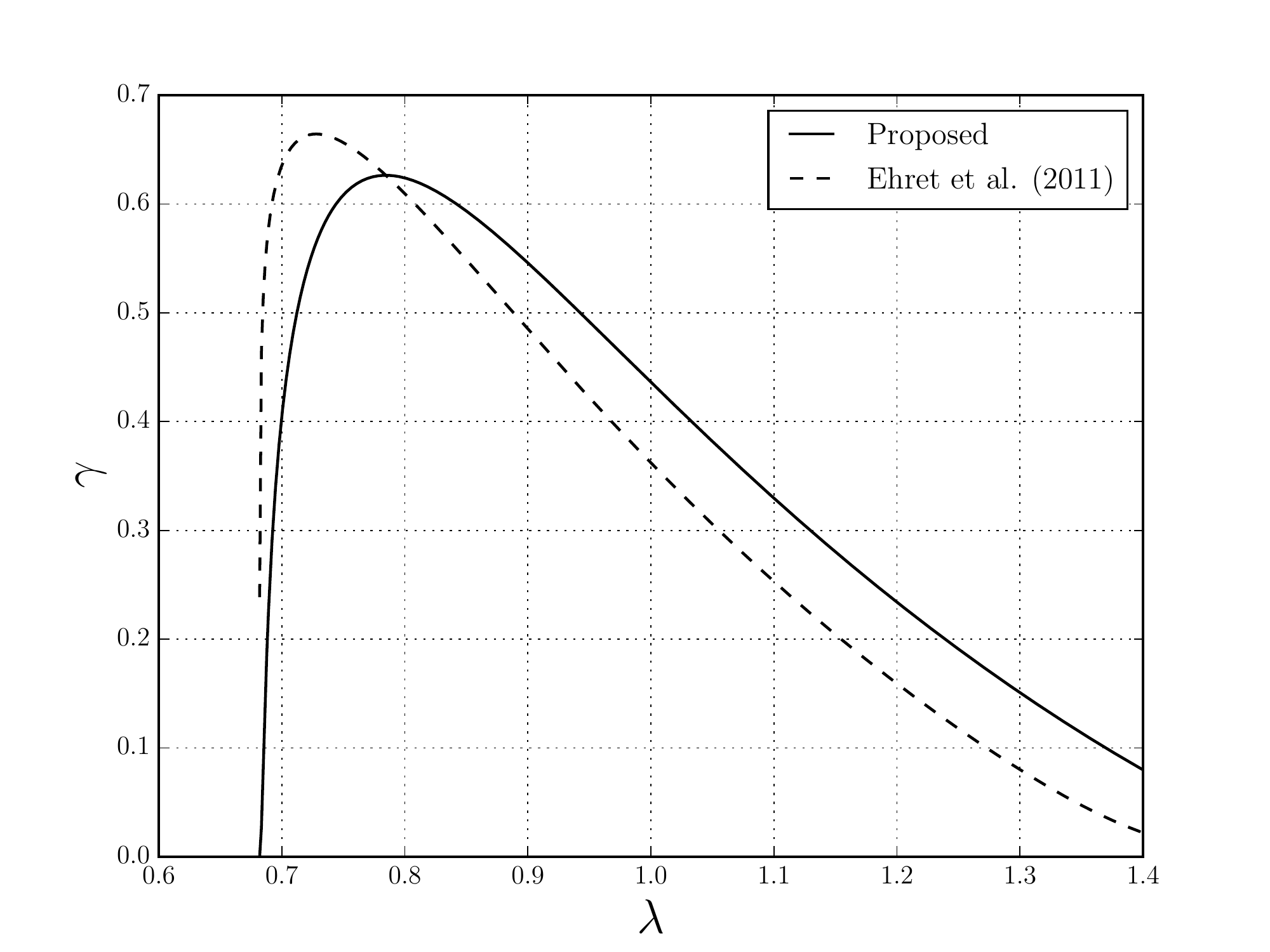}
\caption{Comparison between the behavior of $\gamma(\lambda)$ in
     \eqref{soluzionewagiusta} (solid line) and the one in
     \eqref{waebi} (dashed line).}
\label{figwaconfronto}
\end{figure}

The trend of $\gamma(\lambda)$ is showed in
Figure~\ref{figwaconfronto} in comparison to the one obtained in
Section~\ref{sectebi}.

 \section*{Acknowledgement}

  This work has been supported by the project \emph{Active Ageing and
    Healthy Living} of the Universit\`a Cattolica
  del Sacro Cuore and partially supported by National Group of
  Mathematical Physics (GNFM-INdAM).

\bibliography{../../muscoli}
\bibliographystyle{unsrt}

\end{document}